\pdfoutput=1

\documentclass[twocolumn,showpacs,amssymb,aps,nofootinbib,floatfix,superscriptaddress]{revtex4-1}

\usepackage{epsfig}
\usepackage{hyperref}
\usepackage{comment}

\usepackage{graphicx,color}

\begin{document} 
\hbadness=10000

\title{Transverse momentum-flow correlations in relativistic heavy-ion collisions}

\author{Piotr Bo\.zek}
\email{piotr.bozek@fis.agh.edu.pl}
\affiliation{AGH University of Science and Technology, Faculty of Physics and
Applied Computer Science, al. Mickiewicza 30, 30-059 Krakow, Poland}

\begin{abstract}
The correlation between the  transverse momentum and the
azimuthal  asymmetry of the flow is studied. A correlation coefficient is defined between the average
transverse momentum of hadrons emitted in an event and the square of the elliptic or triangular flow coefficient.
The hydrodynamic model predicts a positive correlation of the transverse momentum with the elliptic flow, 
and almost no correlation with the triangular flow in Pb-Pb collisions at LHC energies. In p-Pb collisions the new correlation 
observable is very sensitive to the mechanism of energy deposition in the first stage of the  collision.
\end{abstract}

\date{\today}



\maketitle


\section{Introduction \label{sec:intro}}

Collective expansion of the fireball in relativistic heavy-ion collisions generates an azimuthally 
asymmetric transverse flow. To a first approximation the collective expansion transforms the azimuthal 
asymmetry of the fireball into the elliptic or triangular flow in the final spectra 
 \cite{Heinz:2013th,*Gale:2013da,*Ollitrault:2010tn}.
An essential issue in the analysis of the hydrodynamic response is the identification 
of the relevant parameters of the initial state governing 
the final response \cite{Teaney:2012ke,*Gardim:2011xv,*Niemi:2012aj,*Bhalerao:2014mua,Mazeliauskas:2015vea}. 
Another important topic recently studied concerns  nonlinearities in the hydrodynamics response
\cite{Teaney:2010vd,*Bhalerao:2013ina,*Qiu:2012uy,*Yan:2015jma,Bhalerao:2014xra,*Noronha-Hostler:2015dbi,Mazeliauskas:2015efa}.

One way to study  nonlinearities in the hydrodynamic response is to measure higher order moments between
flow coefficients \cite{Bhalerao:2014xra}.  Another possibility is to
use the event shape engineering \cite{Schukraft:2012ah}. This technique has been used in a number of 
experimental analyzes
\cite{Esumi:2014wga,*Aad:2015lwa,Adam:2015eta} and theoretical studies
\cite{Bzdak:2011np,*Petersen:2013vca,*Huo:2013qma,*Kopecna:2015fwa}. Experimental results indicate that 
for a subsample of  events with higher elliptic flow the transverse momentum 
spectra get harder \cite{Adam:2015eta}. 

Transverse momentum fluctuations from event to 
event are caused by fluctuations in the initial size of the fireball \cite{Broniowski:2009fm,Bozek:2012fw}.
Correlations between the average transverse flow and the coefficients of aziumthally asymmetric flow could
revel interesting informations both on the correlation in the initial state between the size and
 the eccentricities and on the correlations of the  strength of the hydrodynamic response 
with the flow coefficients. In the following, the correlation coefficient between the 
average transverse flow and the square 
of the elliptic or triangular flow coefficient is proposed as a robust observable to study such effects.

\section{Transverse momentum-flow correlations \label{sec:corrvpt}}

The covariance of any observable ${\cal O}$ with the square of the flow coefficient can be defined
\begin{equation}
cov(v_n\{2\}^2, {\cal O})=\langle \frac{1}{N_{pairs}} \sum_{i \neq k } 
 e^{in\phi_i} e^{-in\phi_k}\left( {\cal O -\langle {\cal O} \rangle } \right)\rangle \ ,
\end{equation}
where the sum is over  pairs of particles not used in the calculation of the observable ${\cal O}$.
The simplest way to achieve it is to use  separate pseudorapidity intervals for the calculation of the flow 
coefficient and ${\cal O}$.
The Pearson coefficient for the correlation between ${\cal O}$ and the flow coefficients is
\begin{equation}
R(v_n\{2\}^2,{\cal O})=\frac{cov(v_n\{2\}^2, {\cal O})}{\sqrt{Var(v_{n}\{2\}^2)Var({\cal O})}}  \ .
\end{equation}
By definition the Pearson coefficient in the range $[-1,1]$.

Specifically, for the average transverse momentum in the event
${\cal O}=[p_\perp]=\frac{1}{N}\sum_{i} p_\perp^i $ we have
\begin{eqnarray}
& & cov(v_n\{2\}^2, [p_\perp] )= \nonumber \\ 
& & \langle \frac{1}{N_{pairs}N} \sum_{i \neq k \neq j  }  e^{in\phi_i} e^{-in\phi_k} 
\left( p_j -\langle [p_\perp] \rangle      \right)        \rangle \ .
\end{eqnarray}
In the following we chose the particles in the sum from three different pseudorapidity intervals A, B, C,
$\eta_i \in [-2.5,-0.75]$, $\eta_k \in [0.75,2.5]$ and $\eta_j \in [-0.5,0.5]$. We have checked that similar results can be obtained 
 using one large interval, but excluding self-correlations. The main reason to use
 three separate pseudorapidity intervals is to reduce  non-flow effects. 
The Pearson correlation coefficient is
\begin{eqnarray}
 & & R(v_n\{2\}^2, [p_\perp] )= \nonumber \\
& & \frac{\langle\frac{1}{N_A N_B} \sum_{i  \in A , k \in B  }  e^{in\phi_i} e^{-in\phi_k}  \frac{1}{N_C} \sum_{j \in C}
\left(  p_j -\langle [p_\perp] \rangle      \right) }{\sqrt{Var\left(\frac{1}{N_A N_B} \sum_{i  \in A , k \in B }  e^{in\phi_i}e^{-in\phi_k}\right)  Var( [p_\perp]_C) } }  \ .
\label{eq:pearson}
\end{eqnarray}
The  Pearson coefficient 
can be calculated from the experimental data to estimate correlations between an observable and the magnitude of
the flow. However, the result depends on  multiplicities in the intervals where the quantities are calculated,
so changing the rapidity intervals or transverse momentum cuts introduces a spurious
 effect due to self-correlations, not related to correlations of the collective quantities.

\section{Self-correlations \label{sec:selfcorr}}

The  Pearson correlation coefficient (Eq. \ref{eq:pearson}), 
normalized  by the variances of $v_n\{2\}^2$ and $[p_\perp]$,  depends strongly on the choice of the 
kinematic range, as  the multiplicities can change. In the presence of  collective flow, one is rather
 interested in extracting the correlation coefficient of the event by event characteristics of the spectra, 
the flow 
coefficient squared and the average transverse momentum. 

The correlation coefficient can  be normalized by the standard deviation of the flow coefficient 
and of the average transverse momentum. For the average  transverse momentum it amounts to use
 the dynamical transverse momentum fluctuations \cite{Adams:2005ka}
\begin{equation}
C_{p_\perp}=\langle \frac{1}{N(N-1)} \sum_{i\neq j} (p_i -\langle[p_\perp]\rangle)(p_j-\langle [p_\perp]\rangle) \rangle  \ .
\label{eq:Cp}
\end{equation}
The variance of the flow coefficient squared can be estimated from
\begin{eqnarray}
 Var(v_{n}^2)_{dyn} &  = & \nonumber \\ & & \langle \frac{1}{N_A(N_A-1) N_B(N_B-1)}  \nonumber \\ 
& &  \sum_{i \neq j  \in A } \sum_{ k \neq l \in B  }  e^{in\phi_i+in \phi_j} e^{-in\phi_k-i n \phi_l}  \rangle \nonumber \\
&  & - \langle \frac{1}{N_A N_B} \sum_{i  \in A , k \in B  }  e^{in\phi_i} e^{-in\phi_k} \rangle^2  \ 
 \end{eqnarray}
or equivalently
\begin{equation}
Var(v_{n}^2)_{dyn}= v_2\{2\}^4 - v_2\{4\}^4 \ .
\end{equation}
The correlation coefficient of the collective parameters in the events is
\begin{equation}
\rho(v_n\{2\}^2, [p_\perp])=\frac{cov(v_n\{2\}^2, [p_\perp])}{\sqrt{Var(v_n^2)_{dyn}C_{p_\perp}} }\ .
\label{eq:corrself}
\end{equation}
The correlation coefficient defined above has two desired features. 
First, the correlation coefficient (\ref{eq:corrself}) is a very good estimate 
of the true correlation of the collective parameters.
This can be checked by comparing  results using realistic finite multiplicity events to results obtained  
by integration of the final spectra. Second,  
the correlation coefficient does depend very weakly on the 
choice of the kinematic range in pseudorapidity. Note that such a small dependence is possible 
 due to non-flow effects or  to the 
pseudorapidity dependence of the flow \cite{Khachatryan:2015oea}.
In the following we call $\rho(v_n\{2\}^2, [p_\perp])$ the flow-transverse momentum correlation coefficient.

Unlike the Pearson coefficient \ref{eq:pearson}  the correlation coefficient \ref{eq:corrself} is not necessarily limited to the range $[-1,1]$. However, if genuine nonstatistical fluctuations 
of $v_n$ and $[p_\perp]$ stem from 
fluctuations of  collective parameters
 of the spectra, the correlation coefficient $\rho(v_n\{2\}^2, [p_\perp])$
measures the correlation between these parameters and should be in the range  $[-1,1]$.

\section{Results from the hydrodynamic model \label{sec:hydro}}

Viscous hydrodynamic model simulations in 3+1-dimensions are performed for Pb-Pb collisions 
at $\sqrt{s}_{NN}=2.76$ TeV and p-Pb collisions at $5.02$ TeV \cite{Bozek:2011ua}.
 The initial conditions are   generated event by event from the Glauber Monte Carlo model.
At the positions of the participant nucleons in the transverse plane $x_i,y_i$ entropy is deposited with a 
Gaussian profile  of width $\sigma=0.4$ fm. The transverse profile is given by a sum of contributions from
all participant nucleons
\begin{eqnarray}
S(x,y) & \propto &  \sum_i \left[(1-\alpha) + N^{\rm coll}_i \alpha \right] e^{ - \frac{(x-x_i)^2+(y-y_i)^2}{2\sigma^2}} \ , \label{eq:ggg}
\end{eqnarray}
where the deposited  strength has a contribution  
$1-\alpha$ ($\alpha=0.15$) times the number of  collisions for nucleon $i$, more
details are given in \cite{Bozek:2012fw}.

\begin{figure}[tb]
\includegraphics[width=0.5 \textwidth]{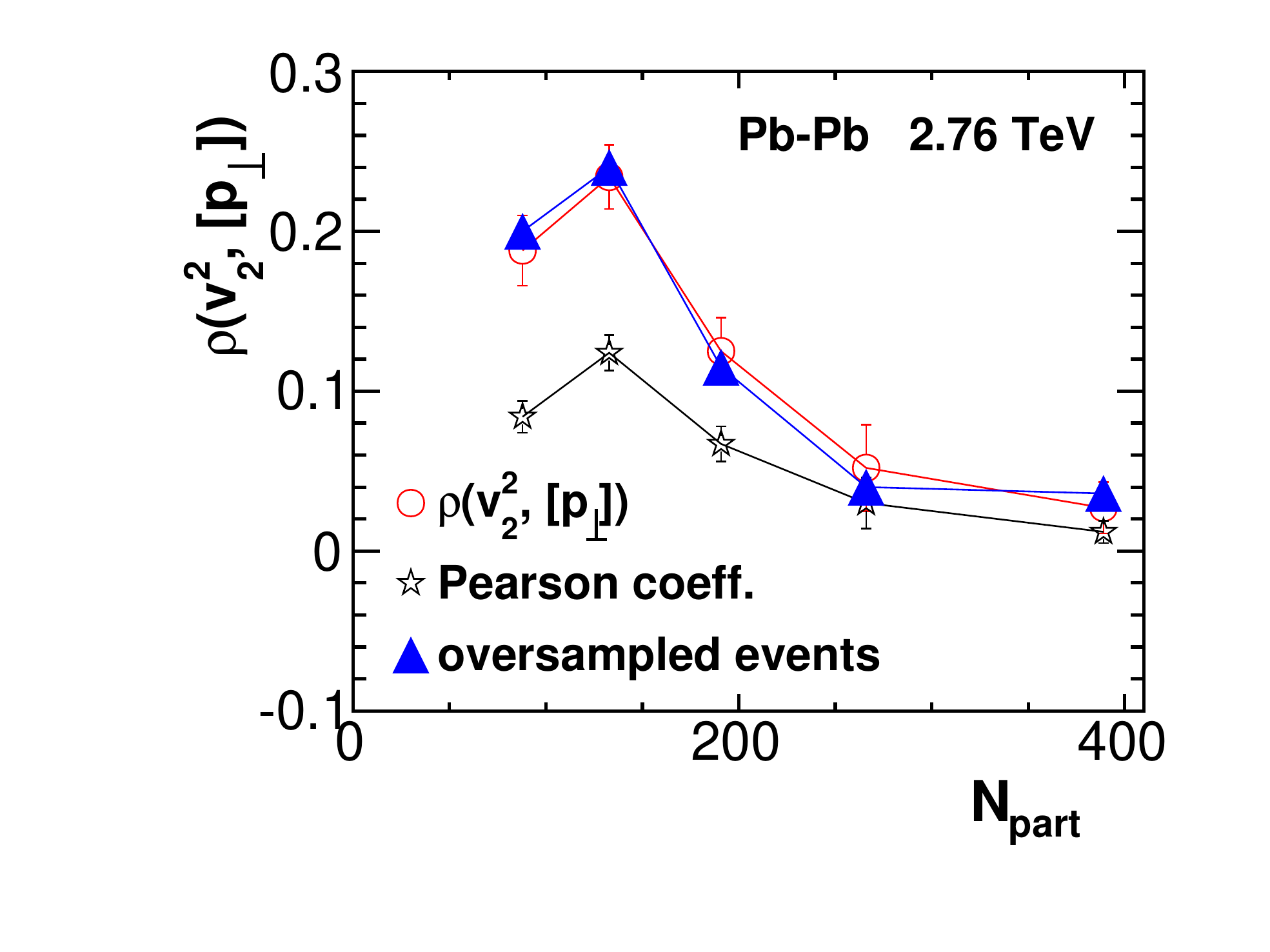}\\ 
\caption{(color online) Correlation coefficient between the elliptic flow coefficient squared $v_2\{2\}^2$
and the average transverse momentum of charged particles in an event for different centralities. The stars
 denote the Pearson coefficient (Eq. \ref{eq:pearson}), the circles denote the correlation coefficient without
 self correlations (Eq. \ref{eq:corrself}) and the triangles denote the correlation coefficient calculated 
from oversampled events.
\label{fig:corrvpt2}}
\end{figure}

\begin{figure}[tb]
\includegraphics[width=0.5 \textwidth]{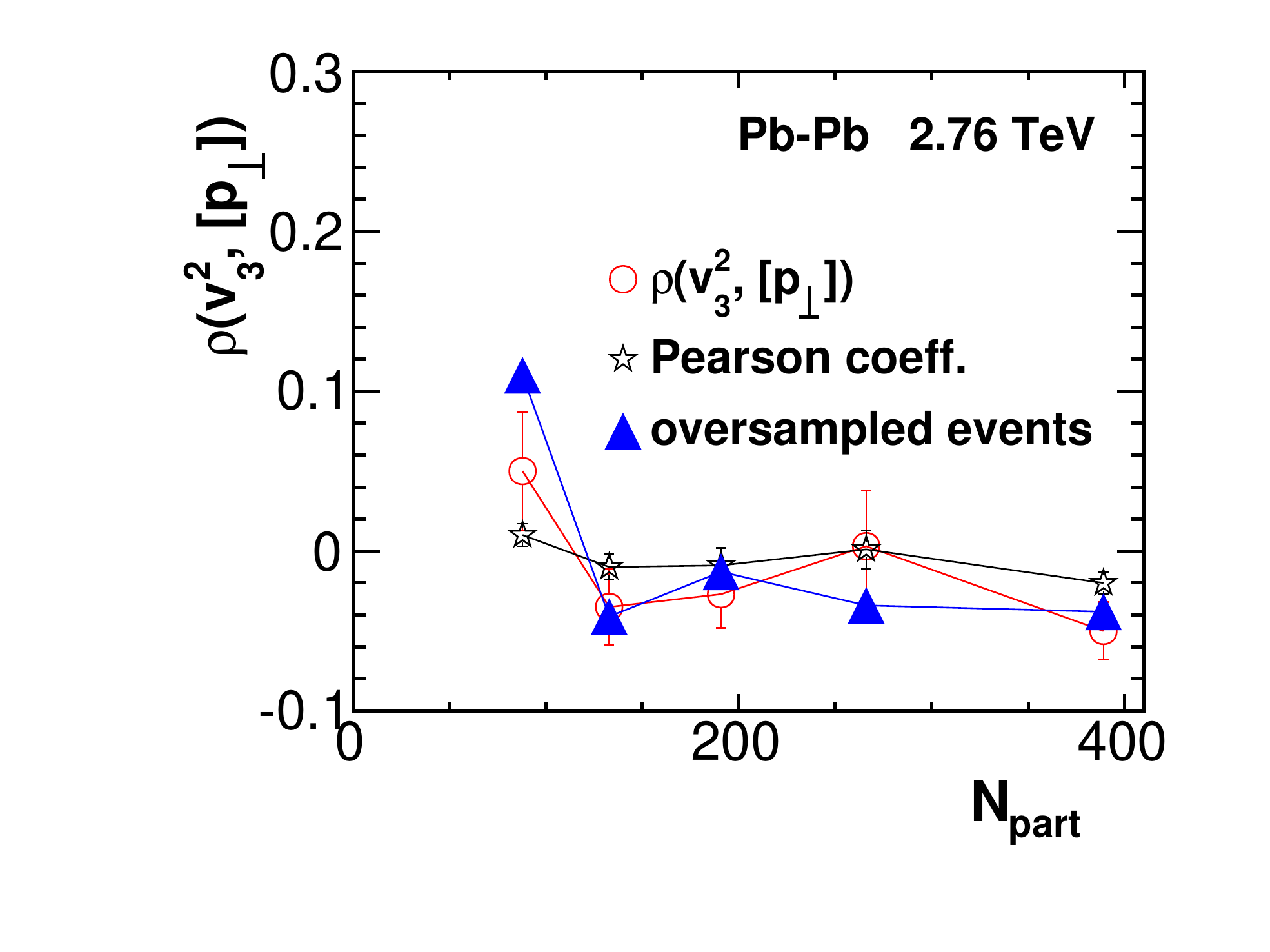}\\ 
\caption{(color online) Same as Fig. \ref{fig:corrvpt2} but for the triangular flow $v_3\{2\}^2$.
\label{fig:corrvpt3}}
\end{figure}

\begin{figure}[tb]
\includegraphics[width=0.5 \textwidth]{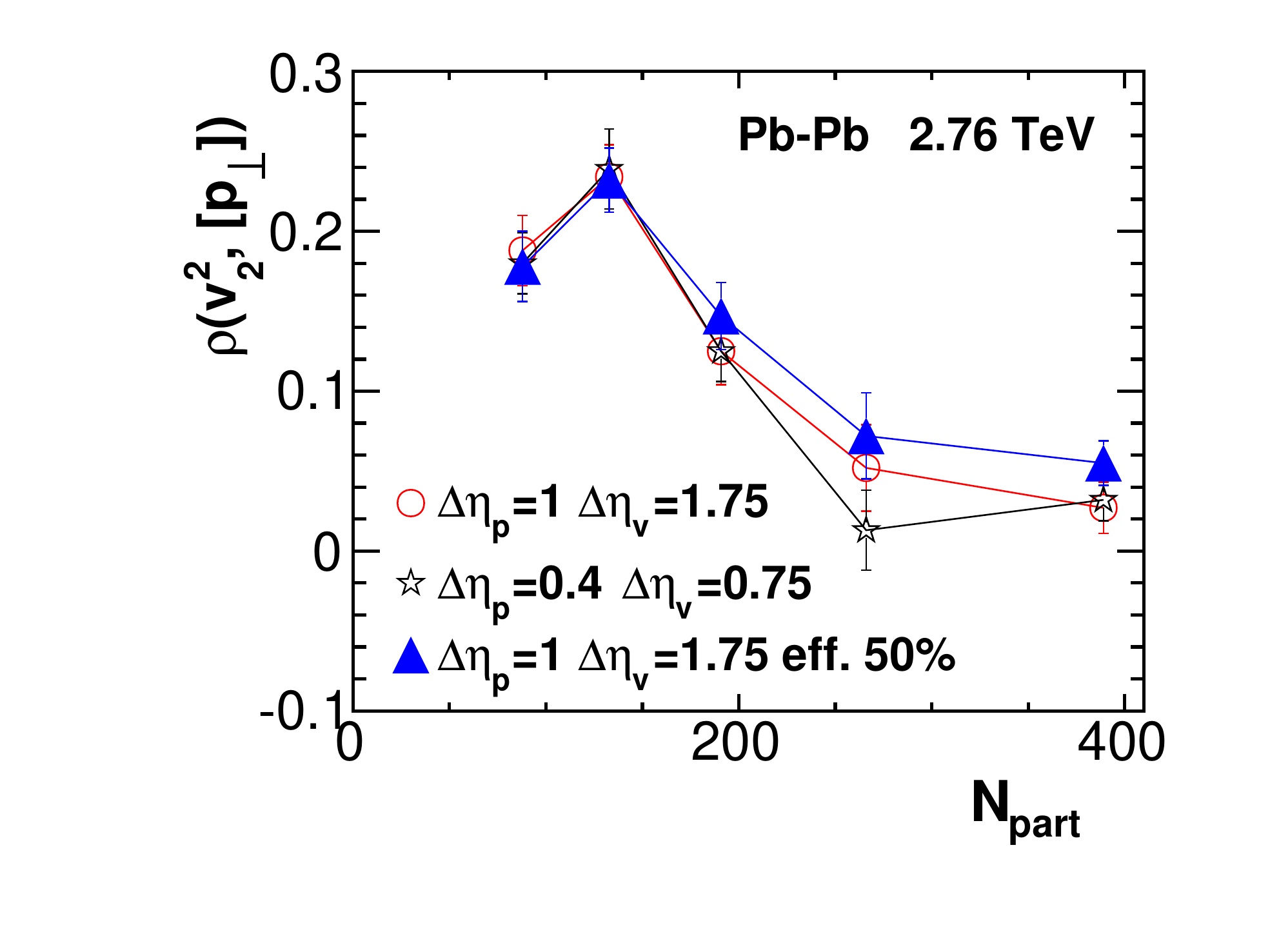}\\ 
\caption{(color online) Correlation coefficient between the elliptic flow coefficient squared $v_2\{2\}^2$
and the average transverse momentum of charged particles in an event for different centralities. The stars
 denote correlation calculated in the range  $|\eta| \in [1.75,-2.5]$ for $v_2^2$, $\eta \in [-0.2,0.2]$ for $[p_\perp]$, the circles and triangles denote the correlation coefficient calculated  with  $|\eta| \in [1.75,-2.5]$ for $v_2^2$, $\eta \in [-0.2,0.2]$ with 100\% and 50\% efficiency respectively.
\label{fig:corrvptacc}}
\end{figure}

In each event, after hydrodynamic evolution, statistical emission of hadrons is performed giving events
 with realistic multiplicities.  The Pearson correlation coefficient (\ref{eq:pearson}) and the correlation 
of the transverse momentum and flow (\ref{eq:corrself}) are calculated in several centrality classes 
from central to mid-peripheral. The centrality classes in the calculation 
are defined by the number of participant nucleons. The Pearson coefficient is always smaller in magnitude than 
the flow-transverse momentum correlation coefficient (Figs. \ref{fig:corrvpt2} and \ref{fig:corrvpt3}).
 It is due to contributions from self-correlation in the denominator of Eq. \ref{eq:pearson}. 
These contributions are important
 for small multiplicities, and get larger for peripheral events or for a  narrow pseudorapidity 
range. The correlation calculated from Eq. \ref{eq:corrself} is a quantity that is defined to be independent on the multiplicity,
 except for small non-flow effects. 
In Figs. \ref{fig:corrvpt2} and \ref{fig:corrvpt3} (triangles) 
are shown the results 
for the flow-transverse momentum correlations obtained by integrating the  spectra in each event. 
Technically, 
these numbers are calculated using oversampled events, where the multiplicity is increased by a factor 100-300 depending
on centrality. As can be observed from the results in Figs. \ref{fig:corrvpt2} and \ref{fig:corrvpt3}, 
the flow-transverse momentum  correlation coefficient  (\ref{eq:corrself}) is very close 
to the result for oversampled events. 
It means that Eq. \ref{eq:corrself} can be used in practice to estimate 
the genuine flow-transverse momentum correlations, without self-correlations  
and with only small non-flow 
contributions. The approximate independence on the pseudorapidity range or efficiency is explicitly shown
in Fig. \ref{fig:corrvptacc}. The results for the correlation coefficient $\rho(v_2\{2\}^2,[p_\perp])$ do
change when the multiplicity changes, due to finite efficiency or different range in pseudorapidity.

The correlation between the elliptic flow and the transverse momentum is positive (Fig. \ref{fig:corrvpt2}), it
is small for central  but
increases for mid-central events. The correlation coefficient reaches $0.25$ for centrality $30$-$40\%$ which indicates
 a significant positive 
correlation. The increase of the mean transverse momentum indicates a stronger
 transverse flow and a stronger collective response to the initial geometry of the source. 
The results are qualitatively consistent with the results of the ALICE Collaboration 
obtained using the event shape engineering technique \cite{Adam:2015eta}.
A stronger transverse push yields a stronger hydrodynamic response of the spectra to the initial azimuthal
 deformation.
 Such an effect is also largely responsible for the observed energy dependence of the integrated elliptic flow 
\cite{Shen:2012vn}. A less important, reverse effect is present in the initial state from the Glauber Monte Carlo
 model. The initial ellipticity  is negatively correlated to the inverse r.m.s radius. Smaller, 
more compact sources give larger transverse momentum, but a smaller deformation. The strength of that negative correlations 
 depends on the width of Gaussian smearing of the deposited density
 from each participant nucleon (Eq. \ref{eq:ggg}). The increase    of the
flow-transverse momentum correlation for mid-central events indicates that it comes  
from  a stronger hydrodynamic response to the large deformation in such events, this is 
consistent with  arguments based on the  principal component analysis of the elliptic and transverse flow
\cite{Mazeliauskas:2015efa}.

The triangular flow shows almost no correlation with the transverse flow (Fig. \ref{fig:corrvpt3}). 
The negative correlation of the initial triangularity with the inverse of the r.m.s radius is stronger than
 for the elliptic flow. Also, the triangular deformation is more sensitive to the initial Gaussian smoothing 
(Eq. \ref{eq:ggg})  than  the elliptic deformation. 
 Unlike for the elliptic flow, 
the magnitude of hydrodynamic transverse push  is not identifiable as a predictor for 
 the triangular flow \cite{Mazeliauskas:2015vea}.

\begin{figure}[tb]
\includegraphics[width=0.5 \textwidth]{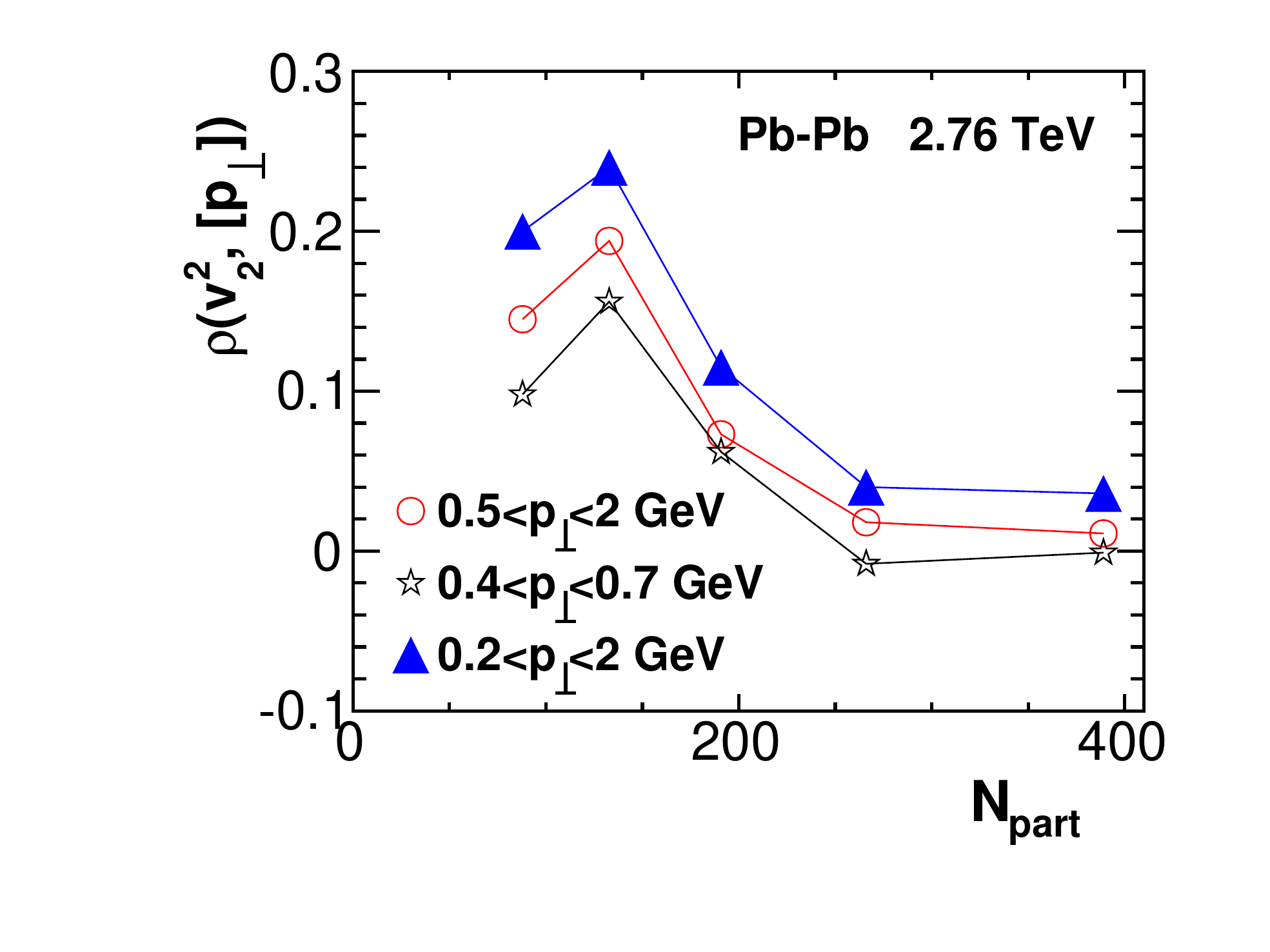}\\ 
\caption{(color online) Correlation  coefficient between the elliptic flow coefficient squared $v_2\{2\}^2$
and the average transverse momentum of charged particles in an event for different centralities. 
The triangles, circles and stars denote the correlation coefficient for the flow calculated in
 the ranges $0.2<p_\perp<2$ Gev, $0.5<p_\perp<2$ Gev, and $0.4<p_\perp<0.7$ Gev  respectively. 
The average transverse momentum of charged particles in the event $[ p_\perp ]$ is calculated 
in the range $0.2<p_\perp<2$ Gev in all the cases.
\label{fig:corrvptrange}}
\end{figure}

The flow-transverse momentum correlation coefficient (Eq. \ref{eq:corrself}) is defined to be independent 
on the  range in pseudorapidity. On the other hand, the elliptic and triangular flows depend 
on transverse momentum. The shift of the integrated elliptic or triangular flow with the change of the average 
transverse momentum depends on the $p_\perp$ range chosen
 to calculate $v_n^2$. In Fig. \ref{fig:corrvptrange}
is compared the correlation coefficient for three different $p_\perp$ ranges used to calculate
 the integrated flow coefficient, $[0.2,2]$ GeV (typical range where predictions of the hydrodynamic model 
are justified), $[0.5,2]$ GeV (preferred range in view of  the efficiency of the ATLAS and CMS detectors at the LHC), 
$[0.4,0.7]$ GeV (range where the average transverse momentum lies). The transverse momentum average is 
calculated for charged hadrons with $0.2< p_\perp < 2.0$ GeV. The calculated flow-transverse momentum 
correlation coefficient depends on the $p_\perp$ integration range.

\begin{figure}[tb]
\includegraphics[width=0.5 \textwidth]{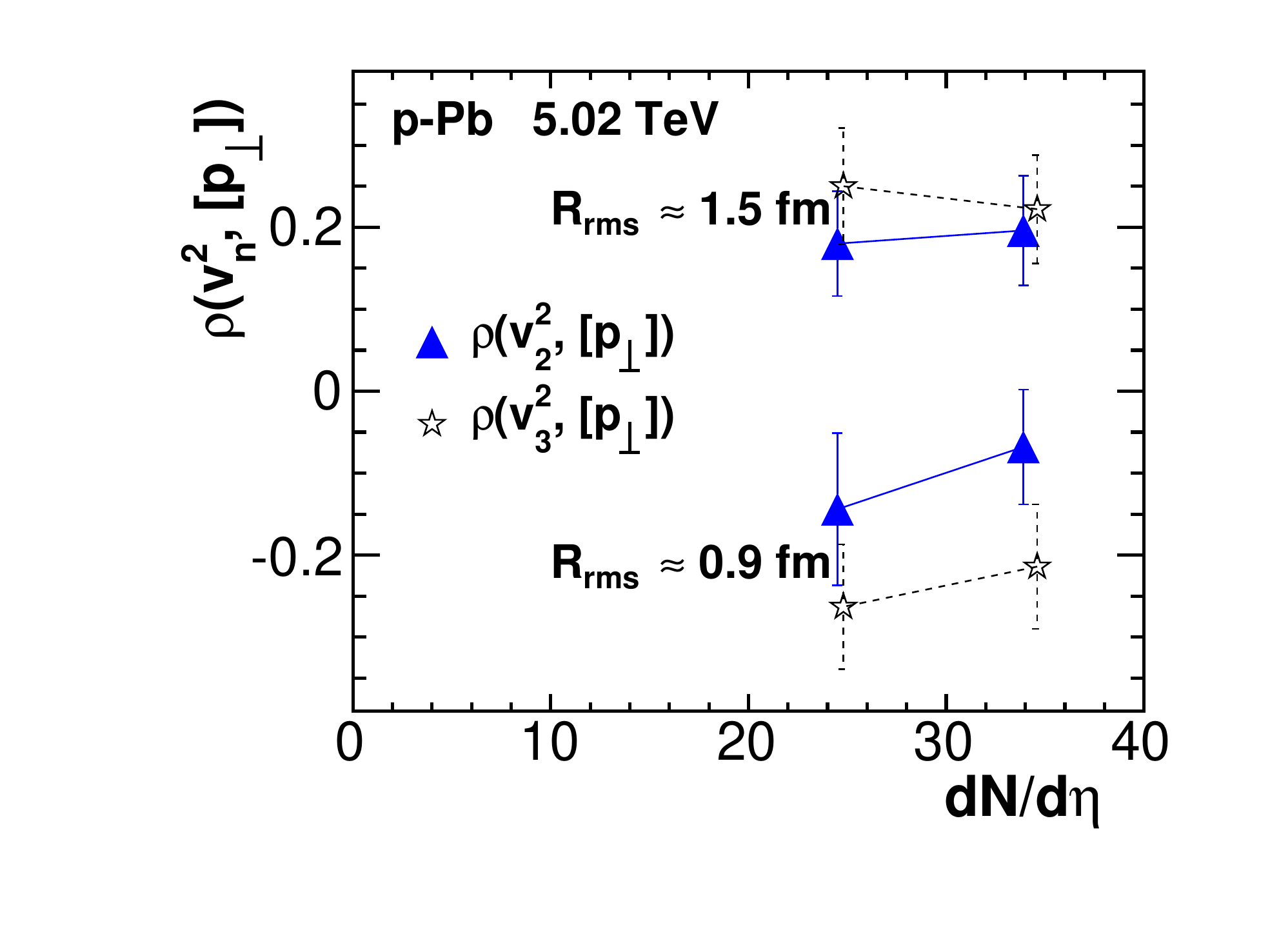}\\ 
\caption{(color online) Correlation  coefficient between the elliptic flow coefficient squared  (triangles) 
or the triangular flow coefficient squared (stars)
and the average transverse momentum for p-Pb collisions at $5.02$TeV for two centralities,
 $0-3$\% and $10-20$\%.  The hydrodynamic evolution is performed for two 
scenarios of energy deposition in the initial fireball, compact source (lower symbols) and standard Glauber model (upper symbols).
\label{fig:corrpPb}}
\end{figure}

The collective flow observed in p-Pb collisions   can be described fairly well using relativistic hydrodynamics \cite{Bozek:2014era}.
The predicted flow depends strongly on  assumptions concerning the initial density fluctuations 
\cite{Bzdak:2013zma}. Two simple scenarios of the entropy deposition in the transverse plane are studied,
 the standard Glauber model, with entropy deposited at the positions of the participant nucleons 
and the compact source scenario, with entropy deposited in between the two colliding
 nucleons \cite{Bozek:2013uha}. For centralities in the range 0-20\% the r.m.s radius of the fireball
in the first scenario is around $1.5$ fm, while in the second case it is much smaller, $0.9$ fm. 

The correlation between final average transverse momentum $[p_\perp]$ and the initial
eccentricites has different sign in the two  scenarios.  For centrality $0$-$3$\%,  we find that 
$\rho([p_\perp],\epsilon_2)=-0.04\pm 0.03$ and $\rho([p_\perp],\epsilon_3)=-0.13\pm 0.04$ 
for the compact source model, while $\rho([p_\perp],\epsilon_2)=0.14\pm 0.03$ 
and $\rho([p_\perp],\epsilon_3)=0.05\pm 0.03$ 
for the standard Glauber model.
 The final flow-transverse momentum correlation is very different in the two scenarios. 
For the larger source it is positive and for the compact source it is negative (Fig. \ref{fig:corrpPb}).
The measured value of the flow-transverse momentum correlation in small systems 
is very sensitive to the mechanism 
of the entropy deposition in the initial state of hydrodynamics. It would be also interesting to check if 
this observable could be used  to distinguish between the hydrodynamic expansion and the 
partonic cascade mechanism \cite{Bzdak:2014dia} of generating flow-like correlations in small systems.

\section{Conclusions}

An observable testing the hydrodynamic response of the particle spectra to the initial eccentricity
is proposed.   It provides a simple quantitative measure of the correlation between the transverse flow
and the coefficients of the azimuthal asymmetry of the spectra. A correlation coefficient can be defined 
between the square of the flow coefficient $v_n^2$ and the average transverse momentum in the event $[p_\perp]$.
Excluding  self-correlations in the calculation of the covariance and the variances, one obtains a good
 estimator of the correlation coefficient of the   average transverse momentum and the flow 
coefficients of the spectra, with no significant non-flow effects. Explicit calculations in the relativistic
 hydrodynamic model show that the flow-transverse momentum correlation
 can be measured for heavy-ion collisions at the LHC. The same is true for collisions at RHIC energies, but 
the smaller pseudorapidity acceptance and smaller  multiplicity would make the interpretation 
more difficult due to non-flow effects.

The azimuthal asymmetry in the spectra of particles emitted in relativistic heavy-ion 
collisions is formed during the collective transverse expansion of the fireball. The strength 
of the response
depends on the gradients of the source density. For smaller sources a stronger transverse flow 
is generated. 
On the other hand, fireballs with a smaller initial size tend to have  smaller eccentricities,
 especially
for the triangular flow. Hydrodynamic model calculations give a significant positive correlation 
between the average transverse flow and the elliptic flow, increasing from central to mid-central 
collisions. 
This is qualitatively consistent with the experimental results using the  event shape engineering 
 \cite{Adam:2015eta} and
the analysis of the nonlinear response in Ref . \cite{Mazeliauskas:2015efa}.
The hydrodynamic model with Glauber model initial condition using a smoothing scale of $0.4$ fm
predicts almost no correlations between the triangular flow and the average transverse momentum.
It would be interesting to check if 
the flow-transverse momentum correlations in peripheral A-A or in p-A collisions 
 could be used as an additional constraint in studies trying to
 estimate the smoothing scale in the initial entropy deposition in the fireball
\cite{Petersen:2010zt,*Renk:2014jja,*Noronha-Hostler:2015coa}.
The sensitivity of the proposed correlation measure to viscosity coefficients of matter in the fireball
is left for further studies. In this context it should be noted that transverse momentum fluctuations 
are sensitive to the effective equation of state \cite{Ollitrault:1991xx}
 and could be sensitive to  the increase 
of bulk viscosity near the critical temperature. 

In small system collisions the magnitude of the transverse push in the expansion is very sensitive to the
duration of the collective dynamics and the size of the initial fireball. Moreover, if the system size
 fluctuates to be small,  the smoothing in the initial entropy deposition yields a stronger reduction of the 
initial eccentricities. The hydrodynamic model gives  very different predictions for the flow-transverse
 momentum correlation in two scenarios, the standard Glauber model and the compact source scenario in p-Pb 
collisions at the LHC.  This observable could be used to probe the mechanism of
  energy  deposition at small scales
in the the first stage of the collision. Finally, it would be interesting to compare the predictions of the 
hydrodynamic  and the cascade AMPT models \cite{Bzdak:2014dia,He:2015hfa} for the
  flow-transverse momentum correlation.

In summary, the paper proposes to study correlations between the flow, or specifically the square of the
 flow coefficient, and other observables, as an alternative to the  event shape engineering technique. 
Hydrodynamic model
 calculations for the correlation of the flow and the transverse momentum demonstrate the practical feasibility 
of the procedure. The flow-transverse momentum correlation could be used to study  fluctuations in the initial stage of the collision.

\begin{acknowledgments}

Research supported by the Polish Ministry of Science and Higher Education (MNiSW), by the National
Science Centre grant DEC-2012/05/B/ST2/02528, as well as by PL-Grid Infrastructure. 
The work has been initiated during 
the program  ``Correlations and Fluctuations in p+A and A+A Collisions'' at the INT, Seattle.

\end{acknowledgments}

\bibliography{../hydr}

\end{document}